

\font\bigbold=cmbx12
\font\eightrm=cmr8
\font\sixrm=cmr6
\font\fiverm=cmr5
\font\eightbf=cmbx8
\font\sixbf=cmbx6
\font\fivebf=cmbx5
\font\eighti=cmmi8  \skewchar\eighti='177
\font\sixi=cmmi6    \skewchar\sixi='177
\font\fivei=cmmi5
\font\eightsy=cmsy8 \skewchar\eightsy='60
\font\sixsy=cmsy6   \skewchar\sixsy='60
\font\fivesy=cmsy5
\font\eightit=cmti8
\font\eightsl=cmsl8
\font\eighttt=cmtt8
\font\tenfrak=eufm10
\font\sevenfrak=eufm7
\font\fivefrak=eufm5
\font\tenbb=msbm10
\font\sevenbb=msbm7
\font\fivebb=msbm5
\font\tensmc=cmcsc10


\newfam\bbfam
\textfont\bbfam=\tenbb
\scriptfont\bbfam=\sevenbb
\scriptscriptfont\bbfam=\fivebb
\def\Bbb{\fam\bbfam}

\newfam\frakfam
\textfont\frakfam=\tenfrak
\scriptfont\frakfam=\sevenfrak
\scriptscriptfont\frakfam=\fivefrak


\def\eightpoint{%
\textfont0=\eightrm   \scriptfont0=\sixrm
\scriptscriptfont0=\fiverm  \def\rm{\fam0\eightrm}%
\textfont1=\eighti   \scriptfont1=\sixi
\scriptscriptfont1=\fivei  \def\oldstyle{\fam1\eighti}%
\textfont2=\eightsy   \scriptfont2=\sixsy
\scriptscriptfont2=\fivesy
\textfont\itfam=\eightit  \def\it{\fam\itfam\eightit}%
\textfont\slfam=\eightsl  \def\sl{\fam\slfam\eightsl}%
\textfont\ttfam=\eighttt  \def\tt{\fam\ttfam\eighttt}%
\textfont\bffam=\eightbf   \scriptfont\bffam=\sixbf
\scriptscriptfont\bffam=\fivebf  \def\bf{\fam\bffam\eightbf}%
\abovedisplayskip=9pt plus 2pt minus 6pt
\belowdisplayskip=\abovedisplayskip
\abovedisplayshortskip=0pt plus 2pt
\belowdisplayshortskip=5pt plus2pt minus 3pt
\smallskipamount=2pt plus 1pt minus 1pt
\medskipamount=4pt plus 2pt minus 2pt
\bigskipamount=9pt plus4pt minus 4pt
\setbox\strutbox=\hbox{\vrule height 7pt depth 2pt width 0pt}%
\normalbaselineskip=9pt \normalbaselines
\rm}


\def\pagewidth#1{\hsize= #1}
\def\pageheight#1{\vsize= #1}
\def\hcorrection#1{\advance\hoffset by #1}
\def\vcorrection#1{\advance\voffset by #1}

\newcount\notenumber  \notenumber=1              
\newif\iftitlepage   \titlepagetrue              
\newtoks\titlepagefoot     \titlepagefoot={\hfil}
\newtoks\otherpagesfoot    \otherpagesfoot={\hfil\tenrm\folio\hfil}
\footline={\iftitlepage\the\titlepagefoot\global\titlepagefalse
           \else\the\otherpagesfoot\fi}

\def\abstract#1{{\parindent=30pt\narrower\noindent\eightpoint\openup
2pt #1\par}}
\def\smc{\tensmc}


\def\note#1{\unskip\footnote{$^{\the\notenumber}$}
{\eightpoint\openup 1pt
#1}\global\advance\notenumber by 1}

\def\frac#1#2{{#1\over#2}}
\def\dfrac#1#2{{\displaystyle{#1\over#2}}}
\def\tfrac#1#2{{\textstyle{#1\over#2}}}
\def\({\left(}
\def\){\right)}
\def\<{\langle}
\def\>{\rangle}
\def\2pd#1#2#3{\frac{\partial^2#1}{\partial#2\partial#3}}

\def\sqr#1#2{{\vcenter{\vbox{\hrule height.#2pt
        \hbox{\vrule width.#2pt height#1pt \kern#1pt
           \vrule width.#2pt}
        \hrule height.#2pt}}}}
\def\square{\mathop{\mathchoice\sqr64\sqr64\sqr{4.2}3\sqr33}}
\def\ni{\noindent}
\def\lqq{\lq\lq}
\def\rqq{\rq\rq}
\def\slash{\!\!\!\!/}


\global\newcount\secno \global\secno=0
\global\newcount\meqno \global\meqno=1
\global\newcount\appno \global\appno=0
\newwrite\eqmac
\def\romappno{\ifcase\appno\or A\or B\or C\or D\or E\or F\or G\or H
\or I\or J\or K\or L\or M\or N\or O\or P\or Q\or R\or S\or T\or U\or
V\or W\or X\or Y\or Z\fi}
\def\eqn#1{
        \ifnum\secno>0
            \eqno(\the\secno.\the\meqno)\xdef#1{\the\secno.\the\meqno}
       \else\ifnum\appno>0
      \eqno({\rm\romappno}.\the\meqno)\xdef#1{{\rm\romappno}.\the\meqno}
        \else
            \eqno(\the\meqno)\xdef#1{\the\meqno}
          \fi
        \fi
\global\advance\meqno by1 }


\global\newcount\refno
\global\refno=1 \newwrite\reffile
\newwrite\refmac
\newlinechar=`\^^J
\def\ref#1#2{\the\refno\nref#1{#2}}
\def\nref#1#2{\xdef#1{\the\refno}
\ifnum\refno=1\immediate\openout\reffile=refs.tmp\fi
\immediate\write\reffile{
     \noexpand\item{[\noexpand#1]\ }#2\noexpand\nobreak.}
     \immediate\write\refmac{\def\noexpand#1{\the\refno}}
   \global\advance\refno by1}
\def\semi{;\hfil\noexpand\break ^^J}
\def\nl{\hfil\noexpand\break ^^J}
\def\refn#1#2{\nref#1{#2}}

\def\up#1{$^{[#1]}$}

\def\cmp#1#2#3{{\it Commun. Math. Phys.} {\bf {#1}} (19{#2}) #3}

\def\ijmp#1#2#3{{\it Int. J. Mod. Phys.} {\bf A{#1}} (19{#2}) #3}
\def\mplA#1#2#3{{\it Mod. Phys. Lett.} {\bf A{#1}} (19{#2}) #3}
\def\pl#1#2#3{{\it Phys. Lett.} {\bf {#1}B} (19{#2}) #3}
\def\np#1#2#3{{\it Nucl. Phys.} {\bf B{#1}} (19{#2}) #3}

\def\prD#1#2#3{{\it Phys. Rev.} {\bf D{#1}} (19{#2}) #3}
\def\prl#1#2#3{{\it Phys. Rev. Lett.} {\bf #1} (19{#2}) #3}

\def\tmp#1#2#3{{\it Theor. Math. Phys.}  (19{#2}) #3}
\def\zpC#1#2#3{{\it Z. Phys.} {\bf C{#1}} (19{#2}) #3}



\def\d{\delta}

\def\phys{{\hbox{\sevenrm phys}}}

\def\L{{\cal L}}

\def\R{{\Bbb R}}

\def\psiphys{\psi_\phys}
\def\psicov{\psi_{\hbox{\sevenrm cov}}}


\pageheight{24cm}
\pagewidth{15.5cm}
\magnification \magstep1
\voffset=8truemm
\baselineskip=16pt
\parskip=5pt plus 1pt minus 1pt


\secno=0

{\eightpoint
\refn\QCD{T. Muta, \lqq Foundations of Quantum Chromodynamics\rqq,
(World Scientific, Singapore, 1987)}
\refn\DIRAC{P.A.M. Dirac, \lqq Principles of Quantum Mechanics\rqq, (OUP,
Oxford, 1958), page 302}
\refn\NEWSYMM{M. Lavelle and D. McMullan, \prl{71}{93}{3758}}
\refn\PROPS{M. Lavelle and D. McMullan, \pl{312}{93}{211}}
\refn\DECAY{S.V. Shabanov, \lqq The Proper Field of Charges and Gauge
Invariant Variables in Electrodynamics\rqq, Dubna preprint,
JINR-E2-92-136}
\refn\KATA{T. Kashiwa and Y. Takahashi, \lqq Gauge Invariance in Quantum
Electrodynamics\rqq,\nl Kyushu and Alberta preprint KYUSHU-HET-14 (1994)}
\refn\JACK{R. Jackiw, in \lqq Current Algebra and Anomalies\rqq,\nl by
S.B. Treiman et al. (World Scientific, Singapore, 1987)}
\refn\GRIB{V.N. Gribov, \np{139}{78}{1};\nl I. Singer, \cmp{60}{78}{7}}
\refn\AXIAL{H. Cheng and E.C. Tsai, \prD{34}{86}{3858};
\nl M. Lavelle and D. McMullan, \zpC{59}{93}{351}}
\refn\GAUGES{M. Lavelle and D. McMullan, \pl{316}{93}{172}}
\refn\CONFINE{See also, M. Lavelle and D. McMullan, \lqq On Quark
Confinement\rqq, Dublin and Mainz preprint\nl DIAS-STP-93-04 and
MZ-TH/93-03, unpublished}
\refn\POLY{A. Polyakov, \lqq Gauge Fields and Strings\rqq, (Harwood,
Chur, 1987)}
\refn\GOJA{J. Goldstone and R. Jackiw, \pl{74}{78}{81}}
\refn\BAGR{V. Baluni and B. Grossman, \pl{78}{78}{226}}
\refn\IZER{A.G. Izergin et al, \tmp{38}{79}{3}}
\refn\DKT{A. Das, M. Kaku and P.K. Townsend, \np{149}{79}{109}}
\refn\TAYL{J.C. Taylor, in proceedings of the workshop \lqq Physical and
Nonstandard Gauges\rqq,\nl Ed.'s P. Gaigg et al (Springer-Verlag, Berlin,
1990)}
\refn\MORSE{G. Dell'Antonio and D. Zwanziger, \cmp{138}{91}{291};
\nl P. van Baal, \np{369}{92}{259}}
\refn\SEIB{We thank David Seibert for correspondence on this point}
\refn\NAKA{See, for example, the discussion in N. Nakanishi and I.
Ojima, \lqq Covariant Operator Formalism of Gauge Theories and
Quantum Gravity\rqq, (World Scientific, Singapore, 1990)}
\refn\SHAB{S.V. Shabanov, \mplA{6}{91}{909}; \pl{255}{91}{398};\nl
L.V.Prokhorov and S.V. Shabanov, \ijmp{7}{92}{7815}}
\refn\POLO{J. Polonyi, \pl{213}{88}{340}}
\refn\BEP{C.M. Bender, T. Eguchi and H. Pagels, \prD{17}{78}{1086}}
\refn\PECC{R.D. Peccei, \prD{17}{78}{1097}}
}
%
%
\rightline {MZ-TH/94-07}
\rightline {PLY-MS-94-06}
\vskip 40pt
\centerline{\bigbold Perturbative and Non-Perturbative Quarks and Gluons}
\vskip 30pt
\centerline{\smc Martin Lavelle{\hbox {$^*$}}{\note{e-mail:
lavelle@vipmza.physik.uni-mainz.de}}
and  David McMullan{\hbox {$^{\dag}$}}{\note{e-mail:
d.mcmullan@plymouth.ac.uk}}}
\vskip 15pt
{\baselineskip 12pt \centerline{\null$^*$Institut f\"ur Physik}
\centerline{Johannes Gutenberg-Universit\"at}
\centerline{Staudingerweg 7, Postfach 3980}
\centerline{D-55099 Mainz, F.R.\thinspace Germany}
\vskip 12pt
\centerline{\null$^{\dag}$School of Mathematics and Statistics}
\centerline{University of Plymouth}
\centerline{Drake Circus, Plymouth, Devon PL4 8AA}
\centerline{U.K.}}
\vskip 7truemm
\vskip 40pt
{\baselineskip=13pt\parindent=0.58in\narrower\ni{\bf Abstract}\quad
BRST invariance supplies a sufficient condition for the
observability of fields. We show that there is a global obstruction
to the observability of quarks and gluons
and argue that they will not become observables at
finite temperature.
We give expressions for quarks and gluons that are, however,
{\it perturbatively\/} BRST invariant, and hence locally observable, up to
order~$g^2$ and~$g$ respectively.
\par}

\vfill\eject
\noindent Quantum Chromodynamics (QCD) is the theory of strong
interactions\up{\QCD}. Its success is based on perturbation theory. The
content of the theory is a non-abelian, $SU(3)$, interaction of quarks
and gluons. Evidence for these particles comes from deep inelastic
scattering. The outstanding problem in QCD is that these particles have not
been observed experimentally. This has led to the confinement hypothesis
that only colour singlet objects are observed in nature. In this letter we
will prove that it is impossible to construct an observable quark or
gluon outside of perturbation theory. Inside perturbation theory, however,
this may be done. We present expressions for such perturbatively
observable quarks and gluons to low orders in the coupling.

The QCD Lagrangian is
$$
\L=-\tfrac14 F^2 +  \bar\psi(i D\slash-m)\psi\,,
\eqn\lag
$$
where $F$ is the field strength constructed out of the non-abelian
(Lie algebra valued) potentials $A$, $D$ is the associated covariant
derivative and
$\psi$ is a fermionic field. This Lagrangian exhibits the
following gauge invariance
$$
\eqalign{
A(x)\to A^U(x) & = U^{-1}(x) A(x) U(x)+ \frac1g U^{-1}(x) d U(x)\,, \cr
\psi(x)\to \psi^U(x) & = U^{-1}(x) \psi(x)\,,
}
\eqn\gauge
$$
where $g$ is the coupling and, for each $x$, $U(x)$ is an element of
SU(3) (or more generally any compact Lie group).
Due to the existence of constraints, which is a direct consequence of
this gauge invariance, observables must be gauge (or, more exactly, BRST)
invariant.
{}From Eq.\thinspace\gauge\  we see that, in particular, the
fermionic fields, $\psi$ and $\bar\psi$, are not observables and thus
cannot be identified with observable quarks. A similar problem for the
electron arises in QED and has been solved by Dirac\up\DIRAC as we now
explain (see also Ref.\thinspace\NEWSYMM).

The physical electron field is given by
$$
\psiphys(x)=\exp\(ig\frac{\partial_i A_i}{\nabla^2}(x)\) \psi(x)\,.
\eqn\electron
$$
This is a spatially non-local field, with $\frac1{\nabla^2}$
being the Green's function for the Laplacian $\nabla^2=\partial_i\partial_i$.
This type of non-locality in the electron field is perfectly
acceptable and reflects the need for an infinite number of soft
photons to deal with the infra-red sector of the theory\up\PROPS.
Indeed, in
contrast to the usual asymptotic
identification of the electron with $\psi$, this BRST invariant expression
has an
electromagnetic charge and creates a Coulomb electric field\up\DIRAC;
justifying its identification with the electron.

In contrast to this, the gauge and BRST invariant expression
$$
\psicov(x)=\exp\(-ig\frac{\partial_\mu A^\mu(x)}{\square^2}\) \psi(x)\,,
\eqn\covon
$$
cannot be identified with a physical field due to the spacetime, and
in particular temporal, non-locality implicit in the
$\frac1{\square^2}$ Green's function.
Being non-local in time obstructs any decomposition into positive and
negative frequency components and hence a particle picture. Indeed
the whole concept of in and out states, as well as time ordering, loses any
meaning for this field. This example, though, does show  the need
for the caveat that BRST invariance is only a {\it sufficient\/} condition
for the observability of a given field.

Another possible candidate for an electron would be
$$
\psi_3(x)=\exp\(-ig\frac{ A_3(x)}{\partial_3}\) \psi(x)\,.
\eqn\axi
$$
Although this is naively gauge invariant, it depends upon an arbitrary
direction which is unattractive in such a physical field. Indeed, a more
detailed investigation\up{\DECAY} shows that this field configuration
is unstable, and decays into the static expression (\electron) proposed
by Dirac.

An at first sight distinctly different ansatz is as follows. Consider
a fermion attached to a Wilson line
$$
\psi_{_{\scriptstyle\Gamma}}(x)=\exp\(ig\int_{-\infty}^x
A_\mu(z)dz^\mu\)\psi(x)\,,
\eqn\string
$$
where $\Gamma$ is any contour from the point $x$ to $-\infty$.
Although this is, by construction, gauge invariant, it is dependent
on the arbitrary line $\Gamma$. A physical electron {\it cannot\/} have
this property. We now require that the contour does not introduce
any temporal non-locality. Then, upon decomposing the spatial components
into the physical, transverse components, $A_i^T$, and the unphysical,
longitudinal component, $A_i^L=\dfrac{\partial_i\partial_j
A_j}{\nabla^2}$, we see that $\psi_{_{\scriptstyle\Gamma}}$ may
be written as
$$
\psi_{_{\scriptstyle\Gamma}}(x)=N_\Gamma(x) \psiphys(x)\,,
\eqn\stringasdirac
$$
where
$$
N_\Gamma(x) =   \exp\(ig\int_{-\infty}^x A_i^T(z)dz^i\)\,.
\eqn\factor
$$
This gauge invariant normalisation factor contains all the contour
dependence and must be removed for the fermion to have any physical meaning.
We thus recover Dirac's physical electron. Indeed it should be noted that
if, as was suggested in Ref.~\KATA, we replace $A_\mu(z)dz^\mu$ by
$A_i^L(z)dz^i$, we obtain just Dirac's electron (since $N_\Gamma$ is then
unity as may be seen from Eq.~\factor).
This concludes our discussion of
QED for the moment.

\bigskip
A sufficient  condition for the non-observability of
quarks would be to show that no contour and gauge independent
generalisation of
Dirac's physical electron can be constructed for the quarks. We
will now demonstrate that this is the case.

Working in a Hamiltonian description\up\JACK of QCD, where
the momentum conjugate to
the potential is denoted by $\Pi(x)$, i.e., such that the fundamental
Poisson brackets are $\{A^a_i(x), \Pi^j_b(y) \}= \d^a_b\d_i^j\d(x-y)$, we see
that if such a field exists it may be written as
$$
\psiphys(x)=h^{-1}(x)\psi(x)\,,
\eqn\hdef
$$
where $h(x)$ is a field dependent element of $SU(3)$. The sought for
gauge invariance of this expression then implies that,
under a gauge transformation, $h$ must transform as
$$
h(x)\to h^U(x)=U^{-1}(x)h(x)\,.
\eqn\htrans
$$
Thus writing
$$
h(x)=\exp\(- v^a(x) T^a  \)\,,
\eqn\hrep
$$
where $T^a$ denotes the (here chosen to be anti-Hermitian)
Gell-Mann matrices,
and expanding the exponential we find that the infinitesimal version of
(\htrans) is
$$
\{v^b(x),G^a(y)\}\approx\d^{ab}\d(x-y)\,,
\eqn\small
$$
where $G^a$ is the infinitesimal generator of gauge transformations
$$
G^a(x)= (D_i\Pi^i)^a(x)+gJ_0^a(x)\,,
\eqn\gauss
$$
and $J_0^a$ is the current density. Note that we are using a weak
equivalence in (\small), i.e., it holds only after
we have set the constraints and $v^a(x)$ to zero. Thus, if such a
construction were possible, we could identify $v^a(x)$ as  possible
gauge fixing functions. (In QED we have seen that it is essentially
the Coulomb gauge.)
We now assume that our fields are chosen so that,
as far as the gauge group is concerned, we can identify the space
time with $\R\times S^3$, where $S^3$ is the spatial
three sphere. As is well known\up\GRIB there is no such
global gauge fixing in QCD (the Gribov ambiguity\note{It is sometimes
argued that algebraic gauges do not suffer from the ambiguity. They are,
however, afflicted with many other diseases, see, e.g., Ref.\thinspace
\AXIAL. In particular these gauge choices are incompatible with
the physical photons\up{\GAUGES}.}).
Hence we deduce that there is no gauge invariant description of a
single quark\up{\CONFINE}.
Of course there are observables in QCD, these correspond to gauge invariant
combinations of the fundamental fields; an example is $\bar\psi\psi$.
We would like to stress that our arguments do not depend upon working
in a particular gauge or indeed upon using a gauge fixing at all:
we merely note that if physical quarks could be defined then they
could be used to construct a globally well defined gauge
fixing. Since this last does not exist, neither do observable quarks.

Quark fields have not been directly observed in experiments and this
has led to the confinement hypothesis, i.e., that they never will be.
We now argue that {\it the non-observability of quark
fields provides an explanation for confinement\/}.
Indeed if we assume to the contrary that quarks are not confined then
they must be directly observed in experiments. In terms of the
structure of QCD, what this would mean is that we can construct a one
quark state (in some asymptotic region). However, we have shown that
it is not possible to construct such a state --- thus we have a
contradiction unless quarks are confined.

We stress that the above is a {\it sufficient\/} condition for
confinement, and is not a necessary one. Indeed,  abelian theories
(for example, compact U(1) in three dimensions\up\POLY) may also
confine due to dynamical effects.

The above simple argument for confinement cannot be considered
watertight --- it skims over many difficult epistemological questions
that plague field theory and indeed quantum mechanics. On a more
immediate level, it fails to convey any scale for confinement. Before
starting  to address this important topic, though, let us briefly
discuss the analogous problem of the observability (or not) of the
gluonic field.

Just as we saw for the fermionic field, the gauge field $A$ cannot be
identified with the physical gluons due to the gauge transformation (\gauge).
However, if we could construct the SU(3)-valued field $h$,
transforming as in (\htrans), then we claim that the physical gluonic
field can be identified with the $h$-transformed potential
$$
A_\phys:=A^h=h^{-1}Ah+\frac1g h^{-1}dh\,.\eqn\gluon
$$
Indeed, under a gauge transformation we have
$$
\eqalign{
A_\phys&\to(A^U)^{U^{-1}h}\cr
&=(U^{-1}h)^{-1}A^U\, U^{-1}h+\frac1g(U^{-1}h)^{-1}d
(U^{-1}h)\cr
&=h^{-1}UA^U\,U^{-1}h+\frac1gh^{-1}(UdU^{-1})h+
\frac1gh^{-1}dh\cr
&=h^{-1}(A+\frac1gd U\,U^{-1})h-
\frac1gh^{-1}(dU\,U^{-1})h+
\frac1gh^{-1}dh\cr
&=A^h\,.
}\eqn\no
$$
So we see that {\it knowledge of $h$ suffices
to specify the physical matter and gauge fields in a gauge theory}.
There is an appealing economy in this description of the physical
fields that is only marred by the global lack of existence of such a
field in the non-abelian theory. However, we can immediately deduce
that in QCD the gluonic field is also not observable. For the abelian
theory we have seen
that we can take
$$
h=\exp\left( -g\frac{\partial_iA_i}{\nabla^2}\right)
\,,
\eqn\hqed
$$
which yields the physical electron (\electron) and from (\gluon) gives
$$
A^i_\phys=\left(\d^{ij}-\frac{\partial^i\partial^j}{\nabla^2}\right) A_j
\,,
\eqn\photons
$$
the two transverse photons. (Indeed, had we used the axial type of
prescription found in (\axi), then the corresponding photon field
would not be transverse. The additional longitudinal  components can be
seen to be responsible for the radiative decay of (\axi) discussed
in Ref.~\DECAY.)
\bigskip
Although, as seen above, no global solution to $(\htrans)$ exists in
a non-abelian theory, we may attempt to find a local solution.
It is important to stress here that local means local in the
Yang-Mills configuration space, which is not directly related to the
spatial locality we want in order to extract a confining scale.
Indeed there are no natural scales in classical QCD, thus we can only
hope to start quantifying our account of confinement if we can
demonstrate that a sensible perturbative expansion of the physical
fields (\electron) and (\gluon) exists in some region of the configuration
space. We are looking for a local solution
in terms of a power series in the coupling~$g$. In this way we
may define {\it perturbatively physical} quarks
and gluons.

The Lagrangian fermion is invariant up to order $g^0$
reflecting the fact that the fermion is a physical particle when the gauge
interaction is switched off. (This is of course not true of  the gauge
fields even in the free photon theory.) When we switch on the gauge
interaction we have to dress the fermion.  A perturbative quark field
which is invariant up to order $g^1$ is seen by inspection to be
$$
\psi_\phys^{g^1}(x)=
\left(1+g\frac{\partial_iA_i^a}{\nabla^2}(x) T^a\right) \psi(x)
\,,
\eqn\psione
$$
which we may suggestively rewrite as
$$
\psi_\phys^{g^1}(x)=
\exp\left(
g\frac{\partial_iA_i^a}{\nabla^2}(x) T^a\right)\psi(x)
+{\cal O}(g^2)
\,,
\eqn\psionea
$$
i.e., to this level in the coupling we reproduce the Dirac electron.
(We must merely replace $A_i$ by $A_i^aT^a$.) Equivalently we have
$$
h^{g^1}(x)=\exp\left( -g\frac{\partial_iA_i^a}{\nabla^2}(x) T^a\right)
+{\cal O}(g^2)
\,.
\eqn\hone
$$
{}From this last equation together with (\gluon) we see that
$$
(A^{g^0}_\phys)^i(x)=\left( \delta^{ij}-
\frac{\partial^i\partial^j}{\nabla^2}\right)A_j(x)\,,
\eqn\gluonone
$$
i.e., to lowest order in the coupling we again obtain the QED result. It
should be noted that due to the factors of $\frac1g$ in (\gluon) and
(\gauge) knowledge of $h$ to any particular power in the coupling only
suffices to specify the perturbatively physical gluons up to one power less
in~$g$.

With the above fields we may perform some lowest order perturbation theory.
The one-loop perturbative quark propagator is just the one loop physical
electron propagator, which was studied in Ref.\thinspace \PROPS, up to an
unimportant colour factor. This is gauge invariant and is the usual
one-loop perturbative fermion propagator in Coulomb gauge. Indeed the
perturbatively physical fields generated by $h^{g^1}$ reduce to the
standard ones in Coulomb gauge as may be seen from (\psionea) and (\gluonone).
\bigskip
We now want to go on to determine $h$ to order $g^2$.
After some algebra one finds that
$$
\eqalign{
h^{g^2}(x)=&1
-g\frac{\partial_iA_i^a}{\nabla^2}(x) T^a
+\frac{g^2}2\left(\frac{\partial_iA_i^a}{\nabla^2}(x) T^a \right)^2
\cr
&+\frac{g^2f^{abc}}2\frac1{\nabla^2}\left(A_j^b\frac1{\nabla^2}\partial_j
\partial_iA_i^c\right)\!\!(x)T^a+\frac{g^2f^{abc}}2
\frac1{\nabla^2}\partial_j\left(A_j^b\frac1{\nabla^2}
\partial_iA_i^c\right)\!\!(x)T^a\,,\cr
=&\exp\Biggl\{\Biggr. \Biggl(\Biggr.-g\frac{\partial_iA_i^a}{\nabla^2}(x)
+\frac{g^2f^{abc}}2\frac1{\nabla^2}\left(A_j^b\frac1{\nabla^2}\partial_j
\partial_iA_i^c\right)\!\!(x)\cr
&\qquad+\frac{g^2f^{abc}}2
\frac1{\nabla^2}\partial_j\left(A_j^b\frac1{\nabla^2}
\partial_iA_i^c\right)\!\!(x)\Biggl.\Biggr)T^a\Biggl.\Biggr\}
+{\cal O}(g^3)\,,\cr}
\eqn\htwo
$$
fulfills this requirement up to terms of order $g^3$.
With this $h^{g^2}$ we find the following perturbatively physical fields:
$$
\eqalign{
\psi_\phys^{g^2}(x)=\Biggl(\Biggr.1
&+g\frac{\partial_iA_i^a}{\nabla^2}(x) T^a
+\frac{g^2}2\left(\frac{\partial_iA_i^a}{\nabla^2}(x) T^a \right)^2
\cr
&-\frac{g^2f^{abc}}2\frac1{\nabla^2}\left(A_j^b\frac1{\nabla^2}\partial_j
\partial_iA_i^c\right)\!\!(x)T^a\cr
&-\frac{g^2f^{abc}}2
\frac1{\nabla^2}\partial_j\left(A_j^b\frac1{\nabla^2}
\partial_iA_i^c\right)\!\!(x)T^a\Biggl.\Biggr)\psi(x)\,,
}\eqn\psitwo
$$
and
$$
\eqalign{
(A_\phys^{g^1})_i(x)=&A_i(x)-\frac{\partial_i\partial_jA_j}{\nabla^2}(x)\cr
&+gf^{abc} T^a\Biggl[\Biggr.\left(\frac{\partial_jA_j^b}{\nabla^2}\right)
\!\!(x)A_i^c(x) +
\frac12\left(\frac{\partial_jA_j^b}{\nabla^2}\right)\!\!(x)
\left(\frac{\partial_i\partial_kA_k^c}{\nabla^2}\right)\!\!(x)
\cr&
+\frac12\frac1{\nabla^2}\partial_i\left(A_j^b\frac1{\nabla^2}\left(
\partial_j \partial_kA_k^c\right)\right)\!\!(x)+
\frac12\frac1{\nabla^2}\partial_i\partial_j\left(A_j^b\frac1{\nabla^2}
\left(\partial_k A_k^c\right)\right)\!\!(x)\Biggr.\Biggl]\,.
}
\eqn\gluontwo
$$
Note that they both reduce to the usual fields in Coulomb gauge and
that the physical gluonic field is transverse to this order, i.e.,
$\partial_i(A_\phys^{g^1})^i=0$. It may be straightforwardly checked that these
fields are indeed invariant under both BRST and gauge transformations up to
terms of order $g^3$ and $g^2$ respectively.

There have been previous attempts\up{\GOJA-\DKT} to find the true degrees of
freedom in pure $SU(2)$ Yang-Mills theory. These solutions have displayed
unpleasant singularities and have not proven to be very tractable.
The suspicion has been
voiced\up{\DKT} that there exists a link between these singularities and the
Gribov ambiguity. From the above results it is clear that this is indeed
the case: due to the Gribov ambiguity it is {\it impossible} to find physical
quarks or gluons and the singularities reflect this.  It is however,
possible to find perturbative solutions and this we have done here. The
field theorists standard tools may now be applied to these solutions.

The results presented here should be used as the basis of a perturbative
investigation of the physical Green's functions of QCD which, as we
have discussed above, is essential if the scale of confinement is to
emerge from this approach (now being identified with the breakdown of
gauge invariance in the locally physical Green's functions). On top
of this, it would be of
interest to see what light they cast upon the infra-red behaviour of the
theory; recall that in QED the infra-red singularities cancel in the
physical Green's functions\up\PROPS. The higher order extension of the
solutions found here should also be investigated. We remark in this
context that problems with the Coulomb
gauge at three loops have been seen by Taylor\up{\TAYL}.

The topological account of the inevitability of the Gribov ambiguity
in non-abelian gauge theories fails to give a feel for the size of
the obstruction. One knows that it is not just the large gauge
transformations that are causing problems. In fact, within the
identity component of the group of all gauge transformations there
are non-contractable loops (for SU(2)), spheres and higher dimensional
objects all conspiring to stop the local observables from becoming global.
However, without any metric information we cannot decide whether such
topological
complications cause  trouble in a large region of the configuration
space or just at a few isolated points (after all, an $n$-sphere
minus one point is topologically trivial). Clearly the need to assess
the extent of the obstruction is related to the space-time confining
scale, and is also important for any consideration of the (local)
stability of the solutions. We have seen that the local expressions
for the quark and gluon, at least to low order in $g$, are closely
connected to Yang-Mills configurations in the Coulomb gauge. Now it
is known\up\MORSE that the Yang-Mills fields which satisfy this
gauge condition can be identified with the critical points of a
naturally defined Morse function constructed out of an
$L^2$-norm on the configuration space. We
would thus expect this norm to be of some use in the analysis of the
extent to which the fields remain physical as they are extended to
the Gribov horizon.

In order to describe QCD at finite temperature and density our assumptions
on the topology of space time must be replaced by $S^1\times S^3$.
This additional complication of the topology will not alter our
arguments, thus we predict that quarks and gluons will not become
observables at high temperature. The scales will change of course and
it is possible that a
large plasma may be generated in which parton-type objects may appear to
move around, analogously to the situation in deep inelastic scattering,
however, individual quarks or gluons will still not be observables\up\SEIB.

Another kinematical account of confinement has been proposed by Kugo
and Ojima\up\NAKA. The connection between their work and ours is
unclear to us, in particular they make no reference to the role of
the Gribov ambiguity. Ideas which appear closer to those advanced here may
be found in Ref.~\SHAB. Also worth noting here is the idea of Polonyi to
link quark confinement to topological obstacles to constructing a gauge
invariant quark propagator\up{\POLO}. That this cannot be done is also, if
on different grounds,  an immediate consequence of our above arguments.
Finally we recall that there have been attempts\up{\BEP,\PECC} to derive
confinement dynamically from the Gribov ambiguity, although these papers
did not concern themselves with the true degrees of freedom of QCD.

\bigskip
To summarise: we have shown that it is impossible to construct physical,
BRST invariant quarks and gluons as a consequence of the Gribov ambiguity.
We interpret this as meaning that quarks and gluons are not observable,
which is one way of stating the confinement hypothesis. We have, however,
been able to construct solutions inside perturbation theory which are
BRST invariant to low orders in the coupling. The one-loop perturbatively
physical quark
propagator was seen to be a close copy of the physical electron propagator
and is hence gauge invariant and infra-red finite.
A detailed perturbative study of the higher order solutions
found here will be presented elsewhere.

\bigskip
\ni{\bf Acknowledgements} MJL thanks the Graduierten Kolleg of Mainz
University for support.
\bigskip

    \vfill\eject
     \immediate\closeout\reffile
  \centerline{{\bf References}}\bigskip\eightpoint\frenchspacing%
  \input refs.tmp\vfill\eject\nonfrenchspacing

\bye